\documentclass[10pt,twocolumn]{article}

\usepackage{graphicx}
\usepackage{amsmath}
\usepackage{amssymb}
\usepackage{authblk}
\usepackage[super,sort&compress,comma]{natbib}
\usepackage{dblfloatfix}

\def\gev{{\, \rm GeV}}
\def\mev{{\, \rm MeV}}
\def\fm{{\, \rm fm}}
\def\eps{\varepsilon}

\let\Im\undefined
\let\Re\undefined
\DeclareMathOperator{\Re}{Re}
\DeclareMathOperator{\Im}{Im}

\usepackage[sc]{mathpazo}
\usepackage[scaled=0.95]{helvet}
\usepackage{courier}

\usepackage[T1]{fontenc}
\usepackage{microtype}
\usepackage{setspace}
\usepackage[margin=1.5cm, top=2cm]{geometry}

\usepackage{color}

\makeatletter
\newcommand{\eprint}[2][]{\@gobble}

\newcommand{\url}[1]{\@gobble}
\makeatother

\title{Strongly coupled quark matter in neutron stars and their mergers}

\author[1,2]{K. Maslov}
\author[2]{R. Rapp}
\author[1,3]{J. Grefa}
\author[3]{V. Dexheimer}
\author[1]{C. Ratti}

\affil[1]{Department of Physics, University of Houston, Houston, TX 77204, USA}
\affil[2]{Cyclotron Institute and Department of Physics and Astronomy, Texas A\&M University, College Station, TX 77843-3366, USA}
\affil[3]{Center for Nuclear Research, Department of Physics, Kent State University, Kent, OH 44242 USA}

\date{}

\begin{document}
\newcommand{\ie}{{\it i.e.}}
\newcommand{\eg}{{\it e.g}}

\twocolumn[
  \begin{@twocolumnfalse}
  \maketitle

  \end{@twocolumnfalse}
]

\textbf{\small 
The discovery of the strongly-coupled quark-gluon plasma (sQGP) in high-energy heavy-ion collisions
has revealed remarkable properties of matter at high temperature,
with transport coefficients close to conjectured bounds from quantum field theory at strong coupling.
The high sQGP collision rates imply very large energy uncertainties
and the melting of quasiparticle structures. Deploying quantum many-body theory based on the self-consistent $T$-matrix approach
for the sQGP at high temperature, we investigate 
its manifestation
at high baryon density and low temperature, as present in neutron stars and their mergers. We constrain the chemical-potential dependence of the quark interaction kernel using first-principles information from Quantum Chromodynamics on baryon-number susceptibilities. Very large collisional widths persist at large density and are found to relegate superconducting phases to rather small temperatures. Instead, a strongly coupled diquark liquid prevails in the thermodynamics under the conditions relevant to neutron-star mergers.
At lower densities, the diquarks 
take over from the single-quark contributions, suggesting a pathway toward hadronization.  Our results are consistent with observational constraints on the equation of state of neutron stars, corroborating the presence of strongly coupled quark matter in the interior of these objects.
}

\section{QCD Matter and Strong Coupling}
Neutron stars are the densest directly observable objects in the Universe, with their innermost parts being compressed to several times nuclear saturation density, $n_0 = 0.16\fm^{-3}$. A fundamental ingredient to describe these systems is the equation of state (EoS)
of Quantum Chromodynamics (QCD). While at vanishing baryon chemical potential, $\mu_B$=0, the EoS is well determined by modern lattice-QCD (lQCD) computations~\cite{Borsanyi:2013bia,HotQCD:2014kol}, its properties (let alone its microscopic structure) at large $\mu_B$ are much less known. From the observational side, precise measurements of pulsar masses~\cite{Fonseca:2021wxt} and of the tidal deformability in the binary neutron-star merger GW170817~\cite{LIGOScientific:2018cki} provide increasingly accurate constraints, while the theoretical predictions are still rather uncertain. At low temperatures, first-principles results for the EoS only exist in the limiting cases of low density, $n \lesssim n_0$, using chiral effective theories~\cite{Hebeler:2013nza,Drischler:2020yad}, and asymptotically high densities, $n \gtrsim 40\, n_0$, using perturbative QCD~\cite{Kurkela:2009gj, Gorda:2023mkk}. The intermediate regime $n_0 \lesssim n \lesssim 10\, n_0$, realized in the core of a neutron star (NS), falls in between them. 

The phase structure in this regime remains largely unknown, e.g., whether there exists a first-order phase transition from baryons (neutrons, protons and their excited states) to deconfined quarks and gluons as the density increases towards the NS core. 
If a deconfined scenario is indeed realized, standard many-body physics predicts that cold dense quark matter features a pairing instability through diquark formation and condensation, thereby forming a color superconductor (CSC)~\cite{Rapp:1997zu,Alford:1998mk}, with pairing gaps of order 100\,MeV or more. This, in turn, would be suggestive of a first-order transition~\cite{Stephanov:1998dy}, although also in this case a ``color-flavor locked" pairing allows for a continuous transition from baryon to quark matter~\cite{Schafer:1998ef}.

This contrasts sharply with the understanding of hot QCD matter from high-energy heavy-ion collisions, where the presence of a strongly coupled liquid has been established~\cite{Shuryak:2008eq}. The extracted transport parameters, such as the ratio of shear viscosity to entropy density~\cite{Heinz:2013th,Bernhard:2019bmu} or the heavy-quark diffusion coefficient~\cite{Dong:2019unq,He:2022ywp} are close to the conjectured values in the strong-coupling limit of quantum field theory~\cite{Kovtun:2004de,Policastro:2002se}, where no readily discernible quasiparticles (QPs) prevail. From the quantum many-body perspective, this implies that the collisional widths in hot QCD matter are so large that energy uncertainties of the would-be QPs become comparable to, or larger, than their masses, i.e., the QPs are no longer well defined.
A quantitative description of cold dense matter should encode these findings. The objective of the present work is to set up and carry out a quantum many-body framework that enables us to transfer the physics of the strongly-coupled quark-gluon plasma (sQGP) into the high-density region. In particular, this framework needs to be able to function in a strong-coupling regime, which requires resummed interactions and a treatment of broad spectral functions beyond the QP approximation.

An example of the required tool is the thermodynamic $T$-matrix approach~\cite{Mannarelli:2005pz,Liu:2017qah}, a non-perturbative many-body method that resums the repeated in-medium scattering of two dressed constituents -- the ladder series -- so that parton dressing, hadron-like bound states, and pairing instabilities are all generated by a single interaction kernel. Very similar diagrammatic methods can be found across various branches of many-body physics: this approach underlies the fluctuation-exchange (FLEX) approximation for strongly correlated electrons~\cite{BICKERS1989206} and the Nozi\`eres--Schmitt-Rink (NSR) description of the BEC-BCS crossover in ultracold atomic systems~\cite{Nozieres:1985zz,Haussmann:2007zz}, and is closely related to the Dirac--Brueckner--Hartree-Fock~\cite{Brockmann:1990cn} method, the NSR~\cite{Tajima:2019saw} treatment of nuclear pairing, and earlier $T$-matrix studies of nuclear matter~\cite{Bozek:1998su,Bozek:1999rv}. Recently, self-consistent real-frequency calculations have become available for ultracold gases, albeit without involving thermodynamic properties~\cite{Enss:2023lau,Dizer:2023mar}. 

In the present work we utilize the thermodynamic $T$-matrix approach to describe the quantum many-body physics of low-temperature quark matter and provide a self-consistent, off-shell calculation with identifiable degrees of freedom that this regime has lacked. We build our approach upon previous work for hot QCD matter, where an in-medium potential, calibrated by lQCD data,
has been deployed to evaluate the properties of heavy quarks and quarkonia~\cite{Riek:2010fk} in connection with a self-consistently calculated EoS~\cite{Liu:2017qah,Tang:2023tkm} of the QGP medium, which also agrees with lQCD results. Parton spectral functions were shown to acquire widths in excess of 500\,MeV and produce transport coefficients compatible with experiment~\cite{Liu:2016ysz}. In the present work, we explore the consequences of the self-consistency for the pairing and EoS of dense quark matter.

\section{Quantum Many-Body Physics for QCD}
Our starting point is the Luttinger-Ward-Baym 
formalism~\cite{Luttinger:1960ua,Baym:1961zz,Baym:1962sx} which provides 
a thermodynamically conserving description of strongly correlated quantum systems. The thermodynamic potential,   
\begin{gather}
	\Omega = \mp \frac{-1}{\beta} {\rm Tr} \Big\{ \ln(-G^{-1}) + \Sigma G \Big\} \pm \Phi[G],
	\label{eq::Omega}
\end{gather}
is formulated as a functional of in-medium particle propagators, $G$, fully dressed by their pertinent self-energies, $\Sigma$, 
and every thermodynamic quantity follows as its derivative.
The convention that the upper (lower) sign applies to bosons (fermions)~\cite{Liu:2017qah} is adopted throughout,
and the trace, ${\rm Tr}$, runs over spin, color, flavor, and energy-momentum (\eg, Matsubara frequencies in the imaginary time formalism).
The first term is the \textit{1-body} contribution, recovering a non-interacting gas for $\Sigma \to 0$, while $\Phi$ is the \textit{Luttinger-Ward functional} (LWF) corresponding to a sum of all 2-particle irreducible diagrams. The fully dressed propagators $G$ obey the Dyson equation
\begin{gather}
	G = G_0 + G_0\,\Sigma\,G .
	\label{eq::dyson}
\end{gather}
where $G_0$ is the bare propagator and the self-energy is self-consistently determined as
\begin{gather}
	\Sigma = \frac{\delta \Phi[G]}{\delta G}.
	\label{eq::selfcons}
\end{gather}
It turns out that the ladder approximation underlying the $T$-matrix resummation,
\begin{gather}
	T = V + V\,G_2\,T \ ,
	\label{eq::tmatrix}
\end{gather}
retains the conserving property of the $\Phi$-derivable formalism, provided it is computed self-consistently with the 1-particle self-energy:
\begin{gather}
	\Sigma = T\,G,
	\label{eq::sigmaT}
\end{gather}
In Eq.~\eqref{eq::tmatrix}, $G_2$ denotes the in-medium 2-particle propagator and $V$ the interaction kernel.
In this setup, $\Phi$ is the sum of particle--particle ladder diagrams with dressed (bold) internal lines (Extended data Fig.~\ref{fig::diagrams}), and accounts, in particular, for dynamically formed 2-body bound states.

The key input to our work is the bare 2-body interaction. In the spirit of the potential approach following previous work~\cite{Riek:2010fk,Liu:2017qah}, and to keep the numerical calculations tractable, we employ a rank-3 separable interaction in $S$- and $P$-waves, where the coupling constant and form-factor cutoffs are adjusted to reproduce the low-lying quarkonium spectrum in vacuum (Extended data Fig.~\ref{fig::quarkonia}), obtained as poles of the corresponding $T$-matrices (Methods).
The resulting interaction is then deployed in the light-parton sector (including relativistic corrections~\cite{Riek:2010fk}) across all $qq$, $q\bar q$, $qg$, and $gg$ color and particle channels, scaled by Casimir factors according to the $SU(3)_c$ representation (Extended Data Table~\ref{tab:casimir}).

With the interaction calibrated, we proceed to evaluate the many-body physics of the QGP at vanishing chemical potential, specifically its EoS. Guided by the procedure carried out for a more sophisticated potential~\cite{Liu:2017qah}, we compute the EoS for temperatures $T \geq T_0 = 0.15\gev$, identifying $T_0$ with the onset of a Debye-type color screening of the interaction (Methods, Eq.~\eqref{eq:formfactors}).
The QGP pressure at $\mu_B=0$ (Fig.~\ref{fig::lattice}\textbf{a}) is fitted by the choice of the in-medium quark mass as a function of the temperature (Extended data Fig.~\ref{fig::ext::mq}). The calculated pressure is dominated by the single-parton contribution at high temperature, but as the temperature decreases, bound states encoded in the LWF $\Phi$ emerge dynamically and supply the main contribution for $T\lesssim0.21\gev$. These correlations include $q\bar q$ meson and $qq$ antitriplet diquark states (Extended data Fig.~\ref{fig::ext::Tmat_mu0}), the latter being the 2-body precursor to baryon formation. Their contribution to the pressure crosses the 1-body contribution at $T\approx0.21\gev$, marking a change in the dominant degrees of freedom of the system. This corroborates that separable interactions can capture the thermodynamic picture of the strongly coupled scenario (SCS)~\cite{Liu:2017qah} for the QGP at $T\lesssim0.25\gev$, corresponding to energy densities comparable to those in the cores of NSs. 
The strong color interaction also produces very large parton widths, with $\Gamma=-2\Im\Sigma$ exceeding $500\mev$ at low momenta. This broadening of spectral functions (Extended data Fig.~\ref{fig::ext::rho_mu0}\textbf{a}) suppresses the 1-body contribution and underlies the liquid-like transport, as in the SCS obtained with the Cornell-type interaction.

\section{Moving to High Density}

\begin{figure}[t!]
	\centering
	\includegraphics[width=\linewidth]{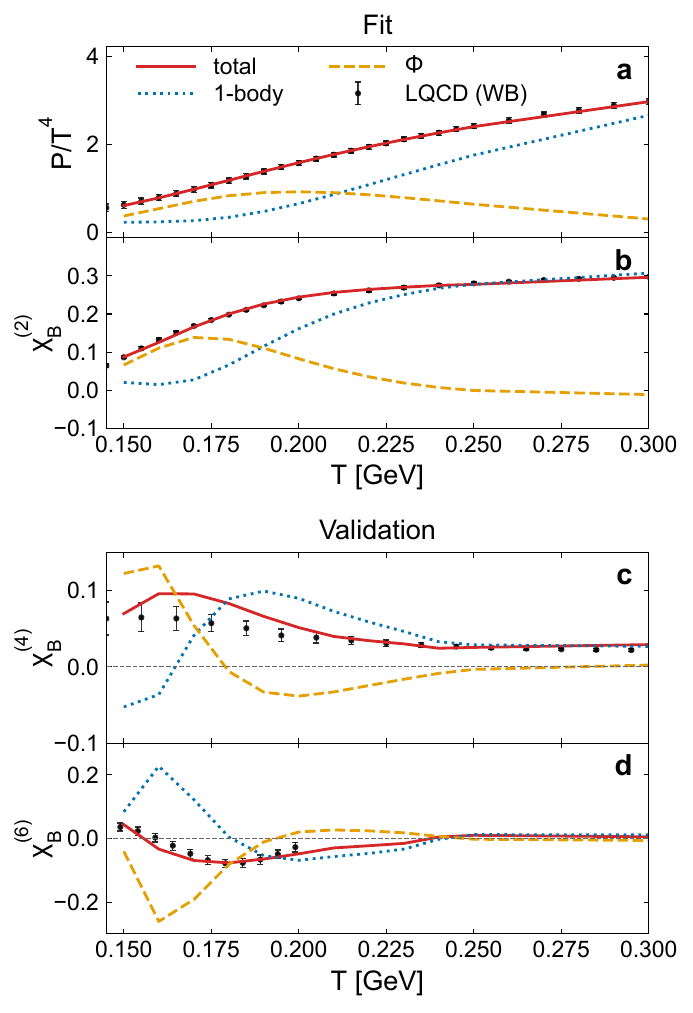}
	\caption{\textbf{Calibration against lQCD thermodynamics at $\mu_B=0$.} \textbf{a}, Pressure $P/T^4$. \textbf{b}--\textbf{d}, Baryon-charge susceptibilities $\chi_B^{(2)}$, $\chi_B^{(4)}$, and $\chi_B^{(6)}$, respectively. Solid lines show the total result, while dotted and dashed lines show its decomposition into the 1-body and Luttinger--Ward $\Phi$ contributions. The quark-mass ansatz is fitted to the data in panels \textbf{a} and \textbf{b}, whereas panels \textbf{c} and \textbf{d} are predictions. Points with error bars are the continuum-extrapolated lQCD data of the Wuppertal--Budapest collaboration~\cite{Borsanyi:2013bia, Bellwied:2015lba, Borsanyi:2023wno}, with error bars denoting the published uncertainties.}
	\label{fig::lattice}
\end{figure}

The extension of our calculation to large $\mu_B$ can be calibrated via the QGP response to increasing baryon charge at $\mu_B=0$, as encoded in the susceptibilities~\cite{Bellwied:2015lba,Borsanyi:2023wno}
\begin{gather}
	\chi_B^{(n)} = \frac{\partial^n (P/T^4)}{\partial (\mu_B/T)^n},
\end{gather}
which have been computed in lQCD.
Figure~\ref{fig::lattice}\textbf{b} shows the baryon-charge susceptibility $\chi_B^{(2)}(T)$, which is calibrated by introducing the chemical-potential dependence into the constituent-quark masses (Extended data Figs.~\ref{fig::ext::mq} and~\ref{fig::beta_02}). Like the QGP pressure, the second-order susceptibility $\chi_B^{(2)}(T)$ includes the 1-body part, corresponding to quark-antiquark asymmetry (Extended data Fig.~\ref{fig::ext::rho_mu0}\textbf{b}). However, the 2-body contribution from the correlations encoded in $\Phi$ is governed by the diquark and antidiquark channels, as mesons do not couple to baryonic charge. Since this contribution is dominant at low $T$, the fit to $\chi_B^{(2)}$ places an upper bound on the diquark coupling strength. 
A consistent treatment of the off-shell diquark spectral properties is therefore essential for capturing this contribution, which, to our knowledge, has not been considered in the literature.

While it is possible to introduce additional parameters for fitting the higher susceptibilities $\chi_B^{(4)}(T)$ and $\chi_B^{(6)}(T)$, it is reassuring to find the $T$-matrix results (Fig.~\ref{fig::lattice}\textbf{c} and \textbf{d}) close to the lQCD data. Our results for $\chi_B^{(4)}$ somewhat overshoot the latter near $T \approx 0.16\,\gev$ but recover the high-temperature behavior for $T \gtrsim 0.2\gev$.
For $\chi_B^{(6)}$, our calculations reproduce the lQCD results well, including the sign change near $T \approx 0.17\,\gev$. While a QP description of $\chi_B^{(2)}$ is available, we note that $\chi_B^{(6)}$ vanishes identically for massless QPs and remains small in the massive case, so the $T$-matrix model provides a qualitative improvement here.

\begin{figure}[t!]
	\centering
	\includegraphics[width=\linewidth]{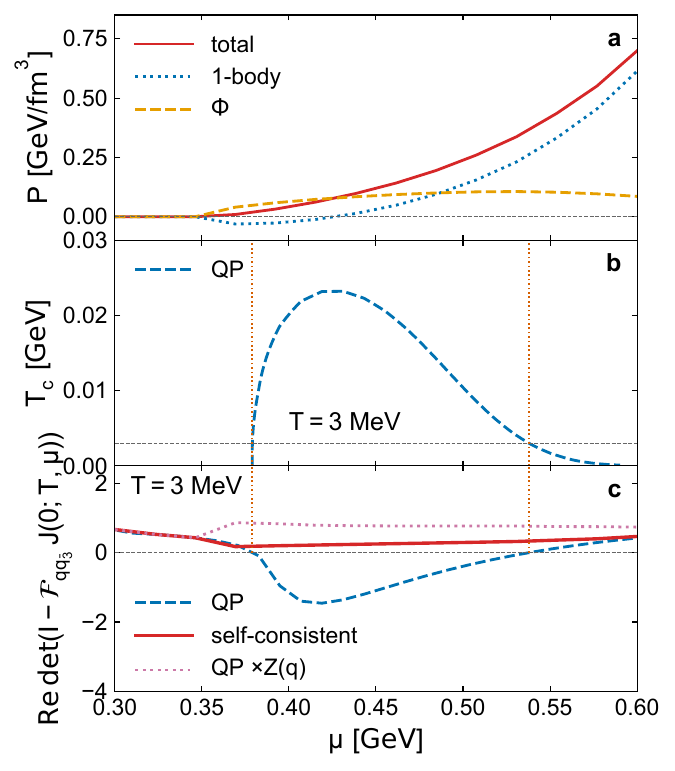}
	\caption{\textbf{Suppression of the pairing instability in the self-consistent calculation.} \textbf{a}, Total pressure (solid) at $T = 3\mev$ and its decomposition into the 1-body (dotted) and correlation (dashed) contributions as functions of $\mu$.
	\textbf{b}, Critical temperature for diquark condensation in the QP approximation as a function of $\mu$. The horizontal dashed line marks $T = 3\mev$, the temperature of the upper and lower panels.
		\textbf{c}, Thouless determinant as a function of chemical potential for the self-consistent calculation (thick solid line), the same QP approximation as in the middle panel (dashed line), and the QP approximation including the QP residue $Z(q)$ (dotted line, see Methods).}
	\label{fig::Pmu}
\end{figure}

\begin{figure*}[t!]
	\centering
	\includegraphics[width=.95\linewidth]{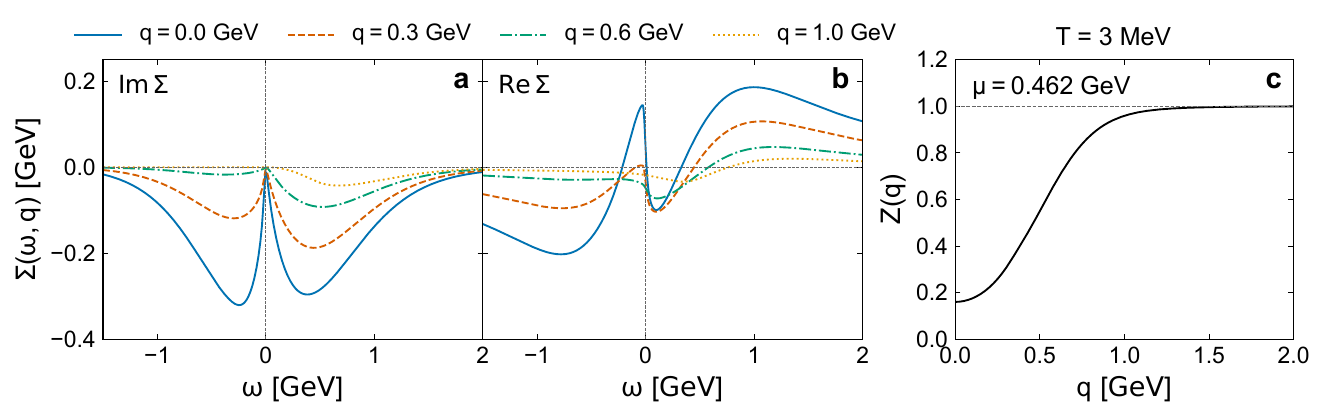}
	\caption{\textbf{Self-energies and the depletion of QP weight.} \textbf{a,} Imaginary and \textbf{b,} Real parts of the retarded quark self-energy $\Sigma(q, \omega)$ at $T = 3\mev$ and $\mu = 0.462\gev$ as functions of the frequency (relative to the Fermi level) for several momenta. The imaginary part exhibits the Fermi-liquid suppression $\Im\Sigma \sim \omega^2$ at the Fermi level $\omega=0$, accompanied by broad structures on the particle and hole sides. These structures induce a steep negative slope of $\Re\Sigma$ at $\omega = 0$, which determines the QP residue. \textbf{c,} QP residue $Z(q)$ (Eq.~\eqref{eq::Z}), showing the reduction to $Z \simeq 0.15$ at low momenta.}
	\label{fig::Zq}
\end{figure*}

We now proceed with the calculation at low $T$ and large $\mu_B$. Within our setup with three degenerate flavors, the model describes charge-neutral beta-equilibrated quark matter and does not require leptons. The resulting pressure as a function of $\mu = \mu_B/3$ (Fig.~\ref{fig::Pmu}\textbf{a}) at $T=3\mev$ exhibits the Silver-Blaze property of QCD~\cite{Fukushima:2010bq}: at low temperature, observables remain at their vacuum values until the chemical potential exceeds a threshold $\mu_{SB}$. In our calculation this property emerges dynamically, as below $\mu_{SB}$ the iteration converges to nearly zero self-energies and the corresponding pressure is close to zero. 
Above threshold, the 1-body contribution from broadened partons starts out negative and crosses zero at $\mu\simeq 0.43 \gev$. The negative contribution, arising from a competition between the first two terms in Eq.~\eqref{eq::Omega}, is a distinct result of our approach, because at those $\mu_B$ the Fermi surface is already present and QP models would predict positive 1-body pressure.
The total pressure stays positive and grows monotonically due to the correlation contribution $\Phi$, and the thermodynamic stability of the matter is maintained by the presence of the non-condensed diquark excitations.

\section{Diquarks and Color Superconductivity}
The convergence of our results at $T= 3\mev$ without the need for deploying the Nambu-Gorkov formalism is somewhat surprising as our diquark binding energy $\sim 200~\mev$ (cf. Extended data Fig.~\ref{fig::ext::Tmat_mu0}) suggests substantial pairing gaps triggering color superconductivity. This requires a more detailed investigation. 
The onset of pairing in the attractive anti-triplet diquark channel can be characterized by the Thouless criterion -- the appearance of a pole in the in-medium diquark $T$-matrix at $E=\vec P=0$ at a critical temperature $T_c$. For a rank-3 separable interaction, this is quantified by the vanishing of the determinant
\begin{gather}
	\Re\det\bigl[\hat 1 - {\cal F}_{qq_{\bar 3}}\,\hat J(E=0, \vec P=0)\bigr] = 0,
	\label{eq::thouless}
\end{gather}
where $\hat J$ is the matrix of dressed two-body loop integrals in potential space (Methods, Eq.~\eqref{eq:ImJ}) and ${\cal F}_{qq_{\bar 3}}$ the Casimir factor of the color-antritriplet channel.
In QP approximation, the $T_c$ for diquark condensation can be obtained as the highest temperature at which the Thouless criterion~\eqref{eq::thouless} is satisfied at a given $\mu$. This corresponds to the BCS approximation, where the Thouless criterion coincides with the linearized gap equation at $T_c$. 
Neglecting the quark self-energy effects in our calculation,
our interaction yields up to $T_c^{\rm max} \simeq 25~\mev$ (see middle panel of Fig.~\ref{fig::Pmu}, i.e., gaps of order 50\,MeV, which is on the lower side of values obtained previously in chiral quark models~\cite{Rapp:1997zu}. We note, however, 
that our diquark coupling is not a free parameter, but has been largely constrained 
by our fit to the $\chi_B^{(2)}$ susceptibility: 
broad diquark states provide a substantial contribution and larger couplings, leading to smaller diquark masses, which would result in an over-prediction of the lQCD data.
The determinant in Eq.~\eqref{eq::thouless} is shown in the lower panel of  Fig.~\ref{fig::Pmu}\textbf{c} as a function of $\mu$ for $T = 3\mev$. The QP result crosses zero at the two values of $\mu$ at which the critical temperature of Fig.~\ref{fig::Pmu}\textbf{b} passes through $T = 3\mev$. However, the full self-consistent Thouless determinant with collisional broadening remains positive, never dropping below $\approx 0.1$, for all $\mu$-values under consideration.
The self-consistent self-energies, shown in the left panel of Fig.~\ref{fig::Zq}, follow the expected scaling~\cite{Luttinger:1960ua, Luttinger:1961zz} $\Im\Sigma(\omega, q) \sim  \omega^2 + O(T^2)$, maintaining the Fermi-liquid behavior at low frequencies, so the absence of the pairing instability cannot be attributed to a breakdown of the Fermi liquid. To understand its origin, we inspect the QP properties at the Fermi surface.

It turns out that the suppression of pairing is connected to the off-shell behavior of the quark self-energy. The width $\Gamma = -2\Im\Sigma$ reaches $\simeq 0.6\gev$ (Fig.~\ref{fig::Zq}\textbf{a}) and extends over energies $|\omega| \lesssim 1\gev$ on both sides (particles and holes) of the Fermi surface. Simultaneously, $|\Im\Sigma|$ remains suppressed in a narrow window around $\omega = 0$, in accordance with the Fermi-liquid $\omega^2$ scaling. By the Kramers-Kronig relation (Methods Eq.~\eqref{eq::methods::KK}), $\Re\Sigma$ is connected to $\Im\Sigma$, and the rapidly varying structures in $\Im\Sigma$ at the Fermi level $\omega=0$ induce a steep negative slope of $\Re\Sigma$ at $\omega = 0$ (Fig.~\ref{fig::Zq}\textbf{b}), which determines the QP residue
\begin{gather}
	Z(q) = \Big(1 - \frac{\partial \Re \Sigma(\omega, q)}{\partial \omega}\Big|_{\omega = 0}\Big)^{-1},
	\label{eq::Z}
\end{gather}
shown in Fig.~\ref{fig::Zq}\textbf{c}. The spectral weight (Extended data Fig.~\ref{fig::ext::lowT}\textbf{a}) is redistributed within $|\omega|<1~\gev$ and partially removed from the Fermi level $\omega=0$.
The residue grows monotonically from $Z(0) \simeq 0.15$ to 1 at $q \simeq 1.5 \gev$. The effect of the decreased QP weight is illustrated by the dotted line in Fig.~\ref{fig::Pmu}\textbf{c}, which corresponds to the QP approximation with inclusion of $Z(q)$, following Methods, Eq.~\eqref{methods::eq::qp_Zq}. Therefore, we can conclude that the decrease of the QP weight is responsible for an increase in the Thouless determinant compared to the BCS case. Similar suppression of $T_c$ due to off-shell dressing was observed in nuclear matter~\cite{Bozek:1999rv, Muther:2005cj}.

\section{Neutron Star Phenomenology}
\begin{figure}[t!]
	\centering
	\includegraphics[width=\linewidth]{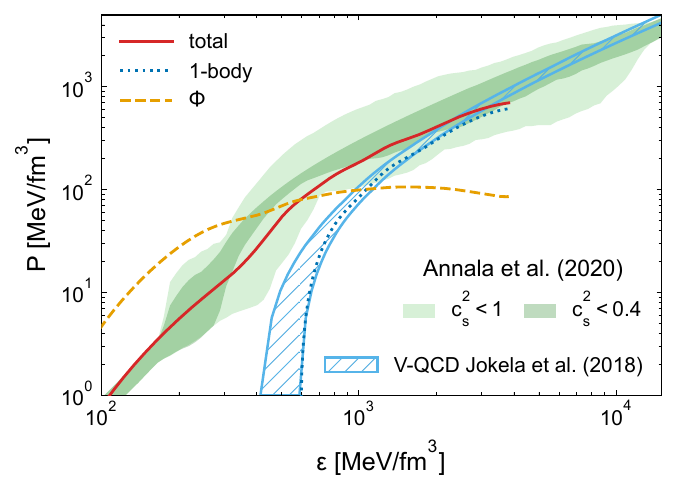}
	\caption{\textbf{EoS of cold quark matter against astrophysical constraints.} Total pressure of charge-neutral beta-equilibrated matter at $T = 3~\mev$ (solid line) and its decomposition into the 1-body (dotted line) and Luttinger--Ward $\Phi$ (dashed line) contributions. The 1-body contribution is negative below $\varepsilon \simeq 0.6\gev/\fm^3$ and is not displayed in the logarithmic scale. The green band shows the model-independent region consistent with causality ($c_s^2 \leq 1$) and astrophysical constraints~\cite{Annala:2019puf}, and the blue hatched band shows the V-QCD calculation~\cite{Jokela:2018ers} with its uncertainty.}
	\label{fig::P_eps}
\end{figure}

Finally, we put our results into the astrophysical context of NSs. Toward that end, we compare our results for the EoS, $P(\varepsilon)$, to another strong-coupling calculation~\cite{Jokela:2018ers} and empirically extracted constraints~\cite{Annala:2019puf} in Fig.~\ref{fig::P_eps}. At low energy densities, up to $\varepsilon \approx 1\gev/\fm^3$, our result is dominated by the contribution of non-condensed diquark fluctuations encoded in the LWF, $\Phi$. The 1-body pressure starts out negative  and crosses zero only at $\varepsilon \simeq 0.6\gev/\fm^3$; it agrees well with the total pressure obtained within the bottom-up  gauge/gravity model in the Veneziano limit (V-QCD)~\cite{Jokela:2018ers}, which has been calibrated to describe the lQCD results for the pressure and $\chi_B^{(2)}$. This suggests the need for a 2-body contribution in that approach, supporting the ongoing efforts to include an explicit diquark degree of freedom~\cite{CruzRojas:2025fzs}.
Our total result for the pressure of quark matter turns out to be surprisingly compatible with the empirical constraints from the existence of massive NSs~\cite{Annala:2019puf}, supporting the presence of deconfined matter in these objects. At low density, our calculation is clearly missing the contributions from colorless hadronic degrees, \ie, nucleons. This can, in principle, be remedied in two ways: (i) the inclusion of colorless 3-body correlations which is expected to raise the pressure~\cite{Tajima:2024qzj}, rendering a non-zero total below 0.3\,GeV/fm$^3$ and better agreement with the constraint for a small speed of sound, and/or (ii) a standard mixed phase construction with a hadronic EoS, which, however, should also be carried out in a strong-coupling framework with large collisional widths, and might yield a rather small latent heat; this remains an open problem for now. 
Either way, the dominant diquark correlations in our calculations can be viewed as precursors to baryon formation.
Note that future calculations of neutrino processes and transport coefficients at temperatures relevant for the warm matter of NS-related transients will not require the deployment of the Nambu-Gorkov formalism for condensed systems but will rather have to rely on off-shell methods with non-trivial particle spectral functions.

\section{Implications}

Our implementation of the strongly coupled QGP as deduced from two decades of research in high-energy heavy-ion physics into the environment typical for the densest objects observable in the universe has revealed some remarkable insights. Utilizing self-consistent quantum many-body theory, whose interactions have been extrapolated to high baryon density with the help of first-principles lQCD data, has enabled a calculation of both thermodynamic and spectral properties of extreme forms of cold quark matter. Very large partonic collisional widths in excess of 0.5 GeV persist, 
but the quantum statistical effects of Pauli blocking of the generated medium create a sharp suppression of the width around the Fermi surface, in line with Fermi liquid theory. Analyticity then implies that the pole strength of the quark excitations near the Fermi surface is pushed to small values, which leads to an unexpected suppression of the BCS instability triggering color-superconducting condensate formation. Nevertheless, strongly correlated yet collisionally broadened diquark configurations continue to play a pivotal role in the system. They form a strongly coupled diquark liquid (sDQL), conceptually akin to the correlation contribution to the density in the description of ultracold atomic gases and nuclear matter~\cite{Ropke:1982ino}.
The very broad single-quark excitations become increasingly suppressed as the density of the system reduces to the extent that the pressure is entirely driven by the sDQL degrees of freedom, which can be viewed as the system evolution toward minimizing the colored degrees of freedom. When comparing  our predictions for the EoS to empirical constraints deduced from neutron stars, we find a good consistency down to energy densities of 0.3 GeV/fm$^3$, which is comparable to the energy densities at the pseudo-critical transition temperature in hot matter at vanishing chemical potential. The good agreement at higher densities therefore corroborates the presence of deconfined matter, although the degrees of freedom are far from QPs.   
In future work, the inclusion of 3-body correlations toward smaller densities is expected to complete the picture to form nucleonic matter and allow for a potentially seamless transition between the partonic and hadronic regimes.  Our microscopic framework also enables a straightforward computation of the transport properties of the sDQL. With markedly different spectral properties compared to, e.g., mean-field approaches, we anticipate new or re-interpretations of pertinent observables such as cooling rates of neutron stars or gravitational waves from their mergers.

\section{Acknowledgements}
This material is based upon work supported by the National Science Foundation under grants No. PHY-2208724, PHY-2116686, PHY-2514763, PHY-2621752 and PHY-2623480, and within the framework of the MUSES collaboration, under Grant No. OAC-2103680. 
The work by RR was supported by the U.S. National Science Foundation under grants no. PHY-2209335 and PHY-2514775.
VD acknowledges support from the U.S. Department of Energy, Office of Science, Nuclear Physics program under Grant DE-SC0024700.
This material is also based upon work supported by the U.S. Department of Energy, Office of Science, Office of Nuclear Physics, under Award Number DE-SC0022023, as well as by the National Aeronautics and Space Agency (NASA) under Award Number 80NSSC24K0767.

{\small
	\bibliographystyle{naturemag}
	\bibliography{biblio.bib}
}

\clearpage
\section*{Methods}

\subsection*{Green's functions}
The central object of the many-body approach used here is the 3D-reduced Green's function for the parton $i = q, \bar q, g$ based on the Thompson scheme
\begin{gather}
	G_i(\omega, p) = \frac{1}{\omega - (\omega_p - \mu_i) - \Sigma_i(\omega, p)},
\end{gather}
where $\mu_{q}=-\mu_{\bar q} = \mu$ and $\mu_g = 0$. In the present formulation the chemical potential enters only through the spectral functions and is absent from the thermal Fermi and Bose distributions. This allows us to easily initialize the iteration from the previously converged point at a different value of $\mu$, without explicit shifts.

For quarks and antiquarks we use the standard relativistic free dispersion relation $\omega_{q,\bar q}(p) = \sqrt{p^2 + m_q^2}$.
For gluons, we enforce two physical polarizations and assume a Gribov-Zwanziger (GZ) \textit{ansatz} for the dispersion relation
\begin{gather}
	\label{eq:gluon-disp}
	\omega_g(p) = \sqrt{m_g^2 + p^2 + \frac{\gamma^4}{p^2}}.
\end{gather}
Our treatment of gluons is closest to the Coulomb gauge, in which a variational Hamiltonian approach to Yang-Mills theory has been shown to lead to the gluon dispersion relation of this form~\cite{Campagnari:2010wc}. We choose the Gribov parameter $\gamma = 0.85\gev$, comparable to the variational estimates, and the mass parameter $m_g = 1.2\gev$ is fixed by fitting the lightest scalar glueball mass $M_{gg} = 1.7\gev$ as obtained in anisotropic-lQCD studies~\cite{Chen:2005mg}. Both $\gamma$ and $m_g$ are kept independent of $T$ and $\mu$. This choice decouples the soft gluons from the rest of the medium and stabilizes our simplified description of the QCD vacuum against the spurious gluon condensation at large temperatures.  Gluons contribute appreciably only at $T \gtrsim 0.3\gev$, above the energy densities reached in NS cores. A consistent treatment of the gauge sector is beyond the scope of the current paper.

\subsection*{T-matrices and self-energies for separable potentials}

The interaction and $T$-matrices $X = (V, T)$ are decomposed into partial waves,
\begin{gather}
	X(\vec q, \vec q\,') = 4\pi \sum_{l} (2l+1)\, X^{(l)}(q, q')\, P_l(\cos\theta_{qq'}),
\end{gather}
with the interaction in each partial wave parameterized in rank-$N$ separable form,
\begin{gather}
	V^{(l)}(q, q') = \sum_{s=1}^N \eta_{s}\, v_s^{(l)}(q)\, v_s^{(l)}(q') \equiv \hat v^{(l)}\, \hat \eta\, \hat v^{(l)} \,.
\end{gather}
The $T$-matrix in color channel $a$ and 2-particle channel $ij$ factorizes as
\begin{gather}
	\langle q | T_a^{(l),i j}(E, P) | q' \rangle = \hat v^{(l)}(q)\; \hat\tau^{(l),ij}_a(E, P)\; \hat v^{(l)}(q'), \\
	\hat \tau^{(l),ij}_a = \frac{{\cal F}_a\, \hat\eta}{\hat 1 - {\cal F}_a\, \hat \eta\, \hat J^{(l),ij} (E, P)} \,,
\end{gather}
where ${\cal F}_a$ is the color Casimir factor (Extended Data Table~\ref{tab:casimir}). The sign matrix $\hat\eta$ is absorbed into $\hat\tau$ and does not appear separately in $\Sigma$, see below.
The loop integral $\hat J^{(l)}$, with $\rho_i = -\frac{1}{\pi}\Im G_i$, reads
\begin{gather}
	\Im J_{s s'}^{(l),ij}(E, P) = 2\int k^2 dk\, \frac{1}{2}\int\limits_{-1}^{1}\! dx\; \frac{m_i}{\eps_i(p_1)} \frac{m_j}{\eps_j(p_2)} \nonumber  \\ \times  v_s^{(l)}(k_{\rm cm})\, v_{s'}^{(l)}(k_{\rm cm}) \int\! d\omega\, \rho_i(\omega, p_1)\,
	\rho_j(E\!-\!\omega, p_2)\,  \nonumber \\  \times \bigl[1 - n_i(\omega) - n_j(E\!-\!\omega)\bigr],
	\label{eq:ImJ}
\end{gather}
where $p_{1,2} = |\vec k \mp \vec P/2|$, $x = \cos\theta_{Pk}$, $n_i$ are the thermal occupations of the two partons (with $-n_{i(j)} \to +n_{i(j)}$ for bosonic statistics of $i(j)$), and $\eps_i(p)$ is the on-shell dispersion relation (Eq.~\eqref{eq:gluon-disp} for GZ gluons). The center-of-mass momentum is
\begin{gather}
	k_{\rm cm}^2 = \frac{\lambda(s,\, m_i^2,\, m_j^2)}{4s},\,\,
	s = \bigl(\eps_i(p_1) + \eps_j(p_2)\bigr)^2 - P^2,
\end{gather}
where $\lambda(a,b,c) = a^2 + b^2 + c^2 - 2ab - 2ac - 2bc$ is the standard K\"all\'en function. The form-factor evaluation at $k_{\rm cm}$ together with the relativistic correction $m_i/\eps_i$ ensures Lorentz invariance of $J$ in vacuum, i.e. $J(E, P) = J(\sqrt{E^2 - P^2}, 0)$. In the QP case used in Fig.~\ref{fig::Pmu}, the quark spectral functions are replaced by their pole parts, $\rho_q(\omega, k) \to Z(k)\, \delta(\omega - \xi_k)$ with $\xi_k = \eps_q(k) - \mu$, where $Z(k) \equiv 1$ corresponds to the BCS case and $Z(k)$ of Eq.~\eqref{eq::Z} to the residue-corrected one. 
At $\vec P = 0$, where $k_{\rm cm} = k$, the loop integral for $qq$ channels simplifies to
\begin{gather}
	\Im J^{(l)}_{ss'}(E, \vec P{=}0) = 2 \int k^2 dk\, \frac{m_q^2}{\eps_q^2(k)}\, v_s^{(l)}(k)\, v_{s'}^{(l)}(k)\, Z^2(k) \nonumber\\
	\times \bigl[1 - 2 n_F(\xi_k)\bigr]\, \delta\bigl(E - 2\xi_k\bigr),
	\label{methods::eq::qp_Zq}
\end{gather}
with the real part entering the determinant of Eq.~\eqref{eq::thouless} restored by the Kramers--Kronig transform defined in Eq.~\eqref{eq::methods::KK}, and $Z(q) = 1$ for the QP case. The QP Silver-Blaze threshold $\mu_{\rm SB}^{\rm QP}$ 
is defined as the value of $\mu$ at which a diquark condensate would have formed in the QP case, $\mu_{\rm SB}^{\rm QP} = m_q(0, \mu_{\rm SB}^{\rm QP}) - E_d^{\rm bind} / 2$,
where $E_d^{\rm bind}$ is the diquark binding energy in the vacuum.
Note that taking the finite width into account shifts the fully self-consistent $\mu_{\rm SB}$ below the QP estimate.
We have checked that increasing the Casimir factor ${\cal F}_{qq_{\bar 3}}$ makes the BCS critical temperature grow in accordance with the expected BCS dependence on the coupling strength, and that even a modest increase of the diquark coupling overshoots the lQCD $\chi_B^{(2)}$.

\subsection*{Self-energy}

Following Ref.~\cite{Liu:2017qah}, the imaginary part of the retarded self-energy of parton $i$ reads
\begin{align}
	&\Im \Sigma_i(\omega, q) = -\frac{1}{\pi} \sum_l (2l\!+\!1) \frac{1}{d_i}\sum_{a,j} d_s^{ij} d_a^{ij} \int p^2 dp \int_{-1}^{1}\! dx\;  \nonumber\\
	&\times \frac{m_i}{\eps_i(q)} \frac{m_j}{\eps_j(p)} \hat v^{(l)}(q_{\rm cm})\; \Im\hat\sigma_a^{(l),ij}(\omega, q, p, x)\; \hat v^{(l)}(q_{\rm cm}),
	\label{eq:ImSigma}
\end{align}
where $x = \cos\theta_{pq}$, $d_i$ is the spin-color degeneracy of parton $i$ ($d_q = 2N_c$, $d_g = 2(N_c^2\!-\!1)$), sum over $j$ is carried over (degenerate) partons, and $d_{s,a}^{ij}$ are the spin and color degeneracies of the two-body channel (Extended Data Table~\ref{tab:casimir}). The reduced self-energy kernel is a matrix in separable space and reads
\begin{align}
	\Im\hat\sigma_a^{(l),ij} = \int\! d\omega'\; &\rho_j(\omega', p)\; \Im\hat\tau_a^{(l),ij}\!\bigl(\omega\!+\!\omega',\, |\vec p\!+\!\vec q|\bigr) \nonumber\\
	&\times\bigl[n_j(\omega') \mp n_{ij}(\omega\!+\!\omega')\bigr],
\end{align}
where $\mp$ refers to the bosonic/fermionic nature of parton $i$, $n_j$ is the thermal distribution of the loop parton, and $n_{ij}$ is the Bose/Fermi distribution appropriate for the two-body state $ij$. The center-of-mass momentum is defined by
\begin{gather}
	s_\Sigma = \bigl(\eps_i(q) + \eps_j(p)\bigr)^2 - (\vec p + \vec q)^2,\nonumber\\
	q_{\rm cm}^2 = \frac{\lambda(s_\Sigma,\, m_i^2,\, m_j^2)}{4\, s_\Sigma}\,.
\end{gather}

For both $f=J, \Sigma$, knowing the imaginary part of an analytic function $f(\omega)$ allows one to restore its real part using the Kramers-Kronig relation
\begin{gather}
	\Re f(\omega) = -\frac{1}{\pi} \mathcal{P} \int \frac{\Im f(\omega') d\omega'}{\omega - \omega'}.
	\label{eq::methods::KK}
\end{gather}

\subsection*{Thermodynamic quantities}

The pressure $P = -\Omega$ is decomposed into two contributions,
\begin{gather}
	P = P_{\rm QP} + P_\Phi\,,
\end{gather}
each summed over parton species $i$ with the spin-color degeneracy $d_i$ ($d_q = 2 N_c$, $d_g = 2(N_c^2 - 1)$), where the species sum is carried over the (degenerate) partons.
The $1$-body QP pressure combines the contribution of dressed single-particle states and the $\Sigma G$ subtraction term:
\begin{align}
	P_{\rm QP} = \sum_i d_i \int\! \frac{d\omega\, q^2 dq}{2\pi^2}\; n_i(\omega)\; \frac{1}{\pi}\Big\{&\delta_i(\omega, q) \nonumber\\
	&- \Im\!\bigl[\Sigma_i\, G_i\bigr]\Big\},
	\label{eq::methods::p1b}
\end{align}
where the overall sign is the same for both statistics, which enter only through the occupations $n_i$, and
\begin{gather}
	\delta_i(\omega, q) = -\Im\ln\!\bigl(-G_i^{-1}(\omega + i0^+)\bigr)
	\label{eq::delta}
\end{gather}
is the phase of the single-particle propagator, which in the QP case rises from $0$ to $\pi$ across the QP peak and reduces to $\pi\,\theta\bigl(\omega - \eps_i(q) + \mu_i\bigr)$ in the free limit.
The Luttinger--Ward functional $\Phi$ resums the interaction contribution beyond mean field using the matrix-logarithm of the $T$-matrix~\cite{Liu:2017qah}:
\begin{gather}
	P_\Phi = \frac{1}{2} \sum_i d_i \int\! \frac{d\omega\, q^2 dq}{2\pi^2}\; n_i(\omega)\; \frac{1}{\pi}\Im\!\bigl[G_i\, \text{Log}\,\Sigma_i\bigr],
\end{gather}
where $\text{Log}\,\Sigma$ is constructed from $\text{Log}\,\hat\tau$ in the same way as $\Sigma$ is from $\hat\tau$ (Eq.~\eqref{eq:ImSigma}). For separable interactions, the matrix logarithm in each color channel reduces to~\cite{Liu:2017qah}
\begin{gather}
	\text{Log}\,\hat\tau_a^{(l)} = -\bigl({\cal F}_a\, \hat\eta\, \hat J^{(l)}\bigr)^{-1}\, \ln\!\bigl(\hat 1 - {\cal F}_a\, \hat\eta\, \hat J^{(l)}\bigr),
\end{gather}
which is a rank-$N$ matrix operation at each $(E, P)$ point. The resulting expression for $P_\Phi$ then simplifies to
\begin{gather}
	P_\Phi = \sum_{ij,\,a,\,l} (2l+1)\, \frac{d_a^{\rm pair}}{2} \int\! \frac{dE\, P^2 dP}{2\pi^3}\; n_a(E)
	 \delta_a^{(l),ij}(E, P),
	\label{eq::methods::Phi_BU}\\
	\delta_a^{(l),ij}(E, P) = -\Im\,\mathrm{Tr}\ln\!\bigl[\hat 1 - {\cal F}_a\, \hat\eta\, \hat J^{(l),ij}(E, P)\bigr],
	\label{eq::methods::delta_2body}
\end{gather}
where the sum runs over the pair species $ij$, their color channels $a$, and partial waves $l$, $d_a^{\rm pair}$ is the total degeneracy of the pair channel, given by the product of the spin-flavor multiplicity of the pair and the color degeneracy $d_a$ of Extended Data Table~\ref{tab:casimir}, the trace is taken in the separable space, and $n_a$ is the Bose (Fermi) distribution for pairs with equal (different) statistics. $P_\Phi$ thus reduces to a single Beth--Uhlenbeck-type phase-space integral~\cite{Ropke:1982ino,Blaschke:2013zaa}, with $\delta_a^{(l),ij}$ being the two-body scattering phase shift derived from the in-medium $T$-matrix.

In our model the bare vertices undergo in-medium screening, which we parameterize as a Debye-like scale. These explicit dependencies on thermodynamic variables induce the rearrangement terms in the thermodynamic quantities~\cite{Liu:2017qah}, structurally similar to well-known rearrangement terms appearing in models of nuclear matter with density-dependent couplings. Thermodynamic consistency is preserved by deriving every bulk quantity---the pressure, the density $n=-\partial\Omega/\partial\mu$, the entropy, and the energy density---from the single thermodynamics potential $\Omega(T, \mu)$, so that the rearrangement terms are included automatically.

\subsection*{Choice of interaction}

We employ a rank-3 separable interaction with the following $s$-wave form factors:
\begin{align}
	v_1^{(0)}(q; T, \mu) &= |G_1| \frac{\Lambda_1^4}{\bigl(\Lambda_1^2 + q^2 + S_1\, \Lambda_1\, \zeta(T,\mu)\bigr)^2}, \nonumber \\
	v_2^{(0)}(q; T, \mu) &= |G_2|\; \frac{e^{-q^2/\Lambda_2^2}}{1 + C_2\, \Lambda_2\, \zeta(T,\mu)\, / T_0^2}\; {\cal R}_2(q), \nonumber \\
	v_3^{(0)}(q) &= |G_3| \frac{\Lambda_3^4}{\bigl(\Lambda_3^2 + (q - q_0)^2\bigr)^2},
	\label{eq::v_s_wave}
\end{align}
where ${\cal R}_2(q) = \sqrt{(\eps_i \eps_j + q^2)/(m_i m_j)}$ is the Breit relativistic correction on $v_2$ (with $\eps_i = \sqrt{m_i^2 + q^2}$), and $v_3$ carries no screening. The coupling sign is $\eta_s = +1$ for all $s$ in the present work.
The $P$-wave form factors are
\begin{align}
	v_1^{(1)}(q) &= \alpha_1^{(p)}\; \frac{q}{\sqrt{\Lambda_1^2 + q^2 + S_1 \Lambda_1 \zeta(T, \mu)}}\; v_1^{(0)}(q) , \nonumber \\
	v_2^{(1)}(q) &= \alpha_2^{(p)}\; \frac{q}{\Lambda_2}\; v_2^{(0)}(q) , \nonumber \\
	v_3^{(1)}(q) &= \alpha_3^{(p)}\; \frac{q}{\Lambda_3}\; v_3^{(0)}(q; q_0 \to q_0^{(p)}).
	\label{eq::v_p_wave}
\end{align}
We limit our parameterization to $S$-wave and $P$-wave channels as the dominant ones. 
The medium-dependent screening variable $\zeta(T, \mu)$ entering $v_1$ and $v_2$ is defined through a nonlinear function of the Debye-like screening scale ${\cal T}$:
\begin{gather}
	\zeta(T, \mu) = \sqrt{{\cal T}(T,\mu) + W^2} - W, \nonumber \\
	{\cal T}(T,\mu) = T_0^2 \max\!\Bigl(0,\;\sqrt{\frac{c_T T^4 + \alpha_1^v\, c_{T\mu}\, \mu^2 T^2 + \alpha_2^v\, c_\mu\, \mu^4}{c_T\, T_0^4}} - 1\Bigr),
	\label{eq:formfactors}
\end{gather}
with the Stefan--Boltzmann energy-density coefficients $c_T = 95\pi^2/60$, $c_{T\mu} = 9/2$, $c_\mu = 9/(4\pi^2)$ for $N_f = N_c = 3$.
The regulator $W$ smoothly interpolates between the linear regime $\zeta \approx {\cal T}/(2W)$ at small ${\cal T}$ and $\zeta \approx \sqrt{{\cal T}}$ at large ${\cal T}$.
All parameter values are listed in Extended Data Table~\ref{tab:params}.

In the present approach, the interaction strength is controlled by the choice of vacuum masses of charm and bottom quarks, which we take as $m_c = 1.85\gev$ and $m_b=5.25\gev$, corresponding to the strongly coupled scenario~\cite{Liu:2017qah}. The parameterization we use provides significant binding energies of $\sim 0.5\gev$ for $c \bar c$ and $1\gev$ for $b\bar b$. The bulk of the binding energy comes from the $v_1$ separable component, while the Gaussian $v_2$ plays a role of color-Coulomb interaction necessary for existence of the excited states, and the position of the resonances allows one to constrain the couplings and form-factors. 
We calibrate the parameters of the interaction to reproduce the spectroscopy of heavy quarkonia, as shown in Fig.~\ref{fig::quarkonia} via the imaginary parts of $T$--matrices in the corresponding channels.
Both partial wave channels reproduce the observed masses within 50\mev. The interaction we use in the light sector differs only by the pertinent relativistic corrections.

The Casimir factors ${\cal F}_a$ and degeneracies $d_a$ for the two-body color channels are listed in Extended Data Table~\ref{tab:casimir}. Attractive channels (${\cal F}_a > 0$) support bound states and resonances near $T_c$; repulsive channels (${\cal F}_a < 0$) contribute to the self-energy but do not form bound states.

\subsection*{Quark mass ansatz}
Since in the present work we do not treat the condensate physics explicitly, the quark masses are treated as fit parameters. 
We also assume that the non-dispersive Hartree and Fock terms are included in the dependence of the quark masses on the temperature using an ansatz $m_q(T, \mu)$.
The lQCD data are available only as a function of $T$ at $\mu = 0$, therefore we need a form that allows an extrapolation to low $T$ and large quark chemical potential $\mu$. For this purpose, we parameterize the quark mass as
\begin{gather}
	m_q(T, \mu) = m_0\frac{T_0^2}{\beta_0(z) T^2 + \beta_2(z) (\frac{\mu}{\pi})^2 + \gamma_2 (\frac{T}{T_0})^2 (\frac{\mu}{\pi})^2},
	\label{eq::m_q}
\end{gather}
where the variable $z = z(T,\mu)$ scales with the energy density of the (massless) system:
\begin{gather}
	z(T, \mu) = \sqrt{\frac{c_T T^4 + \alpha_1 c_{T\mu} T^2 \mu^2 + \alpha_2 c_\mu \mu^4}{c_T T_0^4}} - 1.
\end{gather}
The model is applicable in the domain $z(T, \mu) > 0$, and for $z \leq 0$ the mass saturates at $m_q = m_0$. The explicit values of the fit functions $\beta_{0,2}$ are shown in Extended data, Fig.~\ref{fig::beta_02}, with the intermediate values obtained as cubic spline interpolation.

\subsection*{Self-consistent iteration and numerics}

The coupled Dyson and Bethe-Salpeter equations are solved iteratively: at each step, new self-energies are computed from the current spectral functions $\rho_i$ and loop integrals $J_{ij}$, mixed with the previous iteration, $\Sigma^{(n+1)} = \alpha\, \Sigma^{(n)} + (1-\alpha)\, \Sigma^{\rm new}$, with mixing parameter $\alpha$. Convergence is declared when the largest absolute change of $\Im\Sigma$ over the $(\omega, q)$ grid falls below $10^{-4}\gev$. At $\mu_B = 0$, the iteration converges without damping ($\alpha = 0$), while in cold dense matter we use conservative damping $\alpha = 0.9$, seeding the iteration at each $\mu$ with the converged solution obtained at a larger temperature. At these temperatures the antiquark and gluon excitations are exponentially suppressed, so the cold dense calculations retain only the quark sector with its $qq$ interaction channels.

The numerical grids differ between the two regimes. At $\mu_B = 0$, the energy grid spans $\omega \in [-10, 10]\gev$ with 1601 points ($\delta\omega = 12.5\mev$) and the momentum grid $q \in [0, 5]\gev$ with 200 points. For cold dense matter, the energy grid is strongly refined, $\omega \in [-10, 10]\gev$ with 32001 points ($\delta\omega = 0.625\mev$), with $q \in [0, 5]\gev$ on 200 points. The total pair momentum is sampled in $P \in [0, 5]\gev$ with 200 points. The calculation of frequency convolutions in $\Im J$ is performed using the inverse integration method~\cite{Enss:2023lau}, which resolves the narrow structures at the Fermi surface in the spectral functions and is directly applicable to $\Im J$. 
For the self-energies, the frequency convolutions are evaluated via FFT.

\clearpage

\setcounter{figure}{0}
\renewcommand{\figurename}{Extended Data Fig.}
\setcounter{table}{0}
\renewcommand{\tablename}{Extended Data Table}

\begin{figure*}[t!]
\centering
{\raggedright\LARGE\bfseries Extended data\par}\vspace{1.0em}
\includegraphics[width=.9\linewidth]{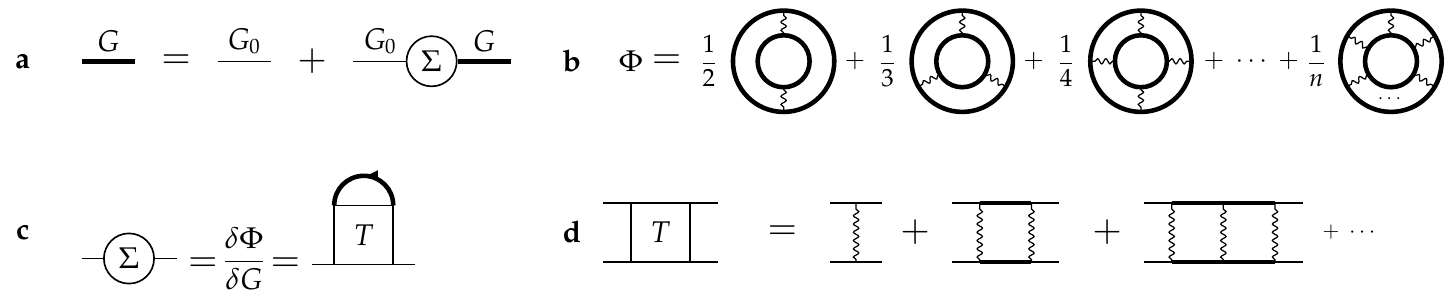}
\caption{\textbf{Diagrammatic content of the self-consistent $T$-matrix approach.}
In all panels, thin lines denote bare propagators, bold lines stand for fully dressed propagators, and wavy lines the bare interaction kernel $V$.
\textbf{a}, Dyson equation relating the dressed propagator $G$ (bold line) to the bare propagator $G_0$ (thin line) and the proper self-energy $\Sigma$.
\textbf{b}, Ladder choice of the Luttinger--Ward functional $\Phi[G]$, with all internal lines dressed. The prefactor $1/n$ accompanies the ring with $n$ interaction lines.
\textbf{c}, Corresponding proper self-energy $\Sigma=\delta\Phi/\delta G$, obtained by cutting the $\Phi$ over a bold line.
\textbf{d}, Bethe--Salpeter equation for the in-medium $T$-matrix with dressed propagators and bare interaction kernel $V$.}
\label{fig::diagrams}
\end{figure*}

\begin{figure}[t!]
	\centering
	\includegraphics[width=\linewidth]{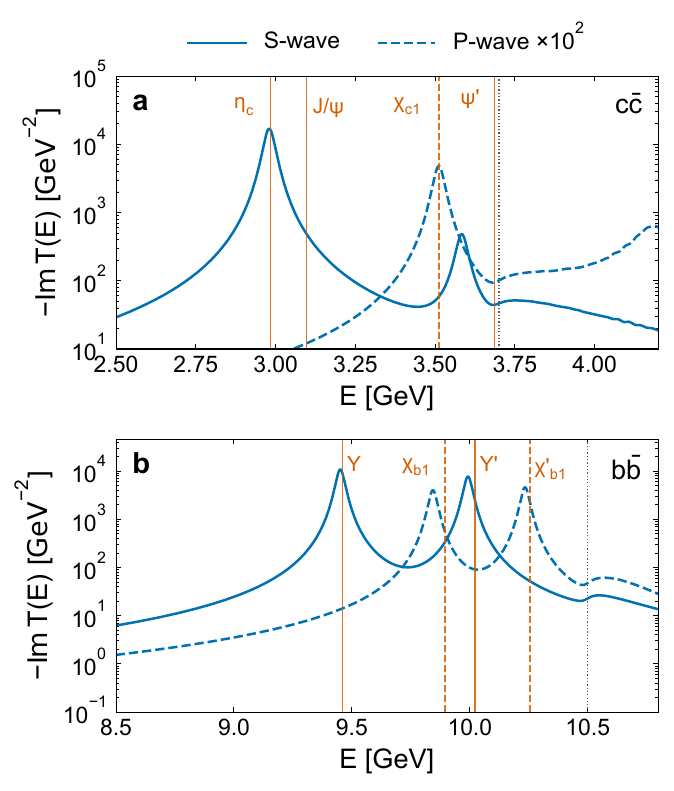}
	\caption{\textbf{Quarkonium calibration in the vacuum}. Imaginary part of the $T$-matrix as a function of the pair energy $E$ at $\vec P=0$ in \textbf{a}, charmonium ($c\bar c$) and \textbf{b}, bottomonium ($b\bar b$) channels with the separable interaction used in this work. Solid (dashed) curves denote the $S$-wave ($P$-wave) channels, the latter scaled by $10^2$ for visibility, with solid (dashed) vertical lines indicating the corresponding PDG values. Dotted vertical lines indicate the two-body threshold $E=2m_Q$.}
	\label{fig::quarkonia}
\end{figure}

\begin{figure}[t!]
	\centering
	\includegraphics[width=\linewidth]{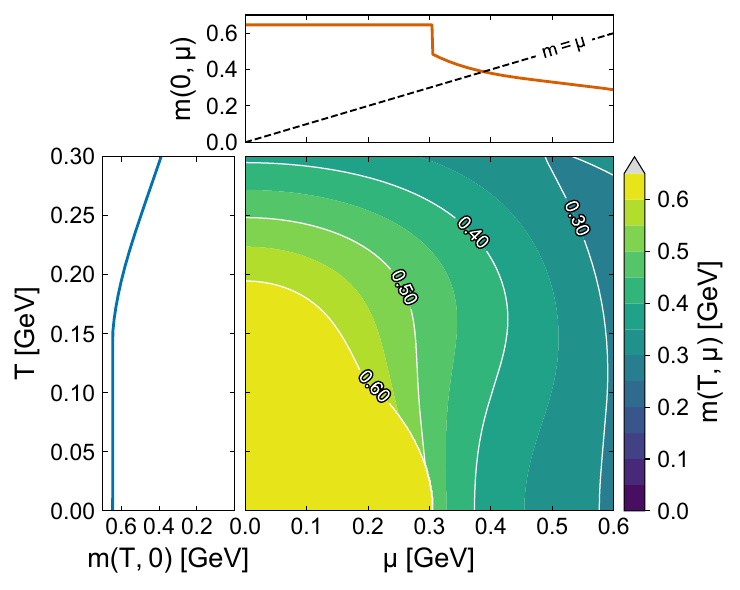}
	\caption{\textbf{Parton mass calibration.} Filled contours of the in-medium quark mass $m(T,\mu)$ (Methods, Eq.~\eqref{eq::m_q}) in the $\mu - T$ plane, with labelled contour lines at $0.30$, $0.40$, $0.50$ and $0.60\gev$. Side panels show the two slices of the same surface, $m(0,\mu)$ along the top and $m(T, 0)$ along the left. The dashed line shows the $m = \mu$ condition that determines the QP Silver-Blaze onset if the diquarks are not considered.}
	\label{fig::ext::mq}
\end{figure}

\begin{figure}[t!]
	\includegraphics[width=\linewidth]{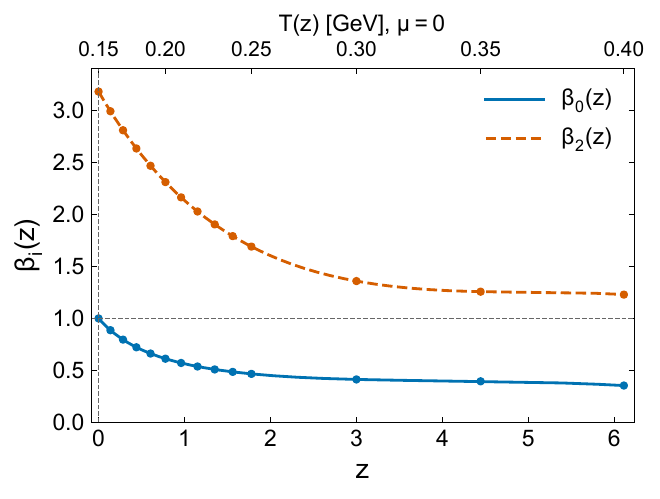}
	\caption{\textbf{Fit functions determining the mass ansatz.} The dimensionless functions $\beta_0(z)$ and $\beta_2(z)$ (Methods Eq.\eqref{eq::m_q}) as functions of the dimensionless variable $z$. The upper axis shows the corresponding values of temperature at $\mu=0$.}
	\label{fig::beta_02}
\end{figure}

\begin{figure*}
	\includegraphics[width=.45\linewidth]{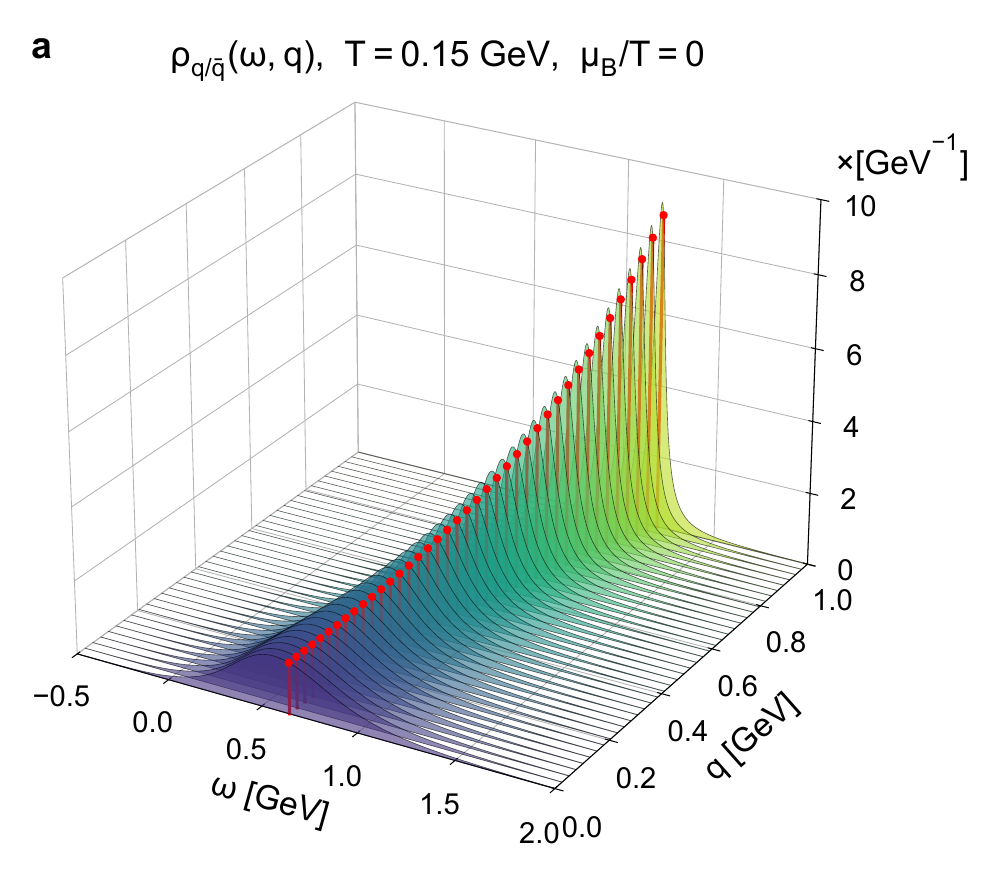}
	\includegraphics[width=.45\linewidth]{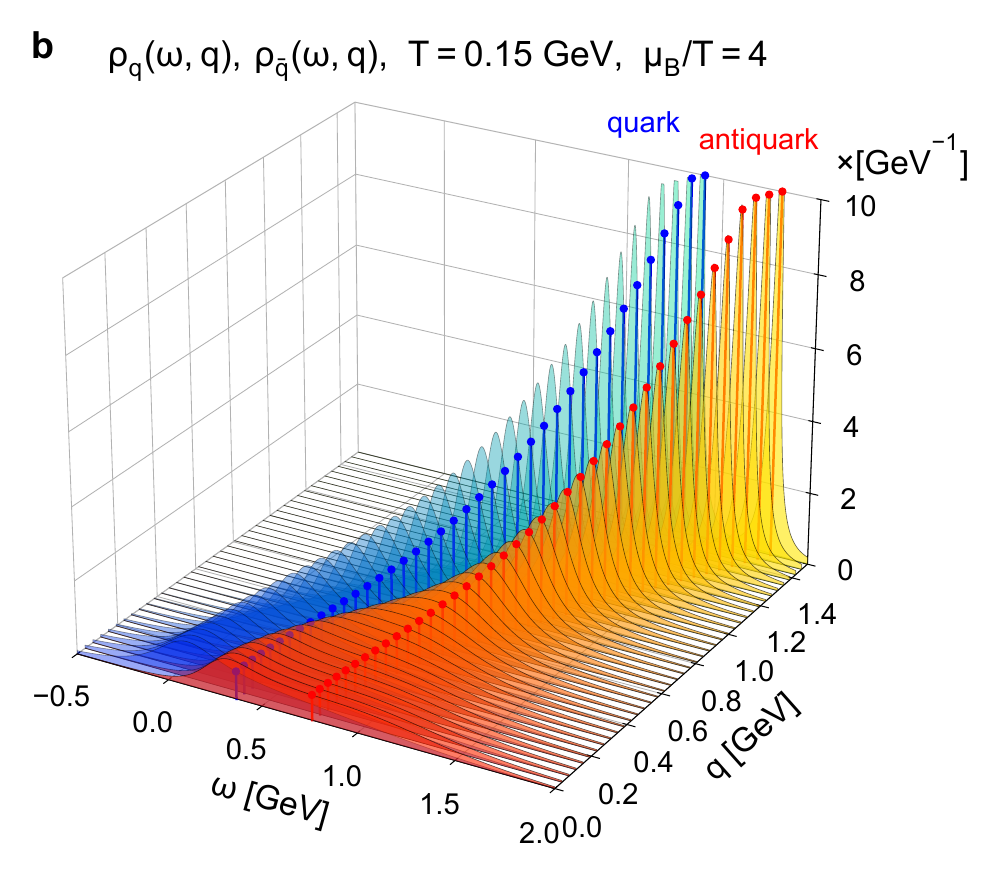}
	\caption{\textbf{Self-consistent parton spectral functions in the QGP.} Spectral functions $\rho(\omega, q)$ at $T = 0.15\gev$, drawn as filled ridges, one frequency slice per momentum $q$. \textbf{a,} Quarks and antiquarks at $\mu_B=0$, where the two are degenerate. \textbf{b,} Quarks and antiquarks at $\mu_B/T=4$, where the chemical potential splits them. Lines ending with dots mark the corresponding QP dispersion relations, the dot being placed at the value of $\rho$ at that energy.}
	\label{fig::ext::rho_mu0}
\end{figure*}

\begin{figure*}
	\includegraphics[width=.45\linewidth]{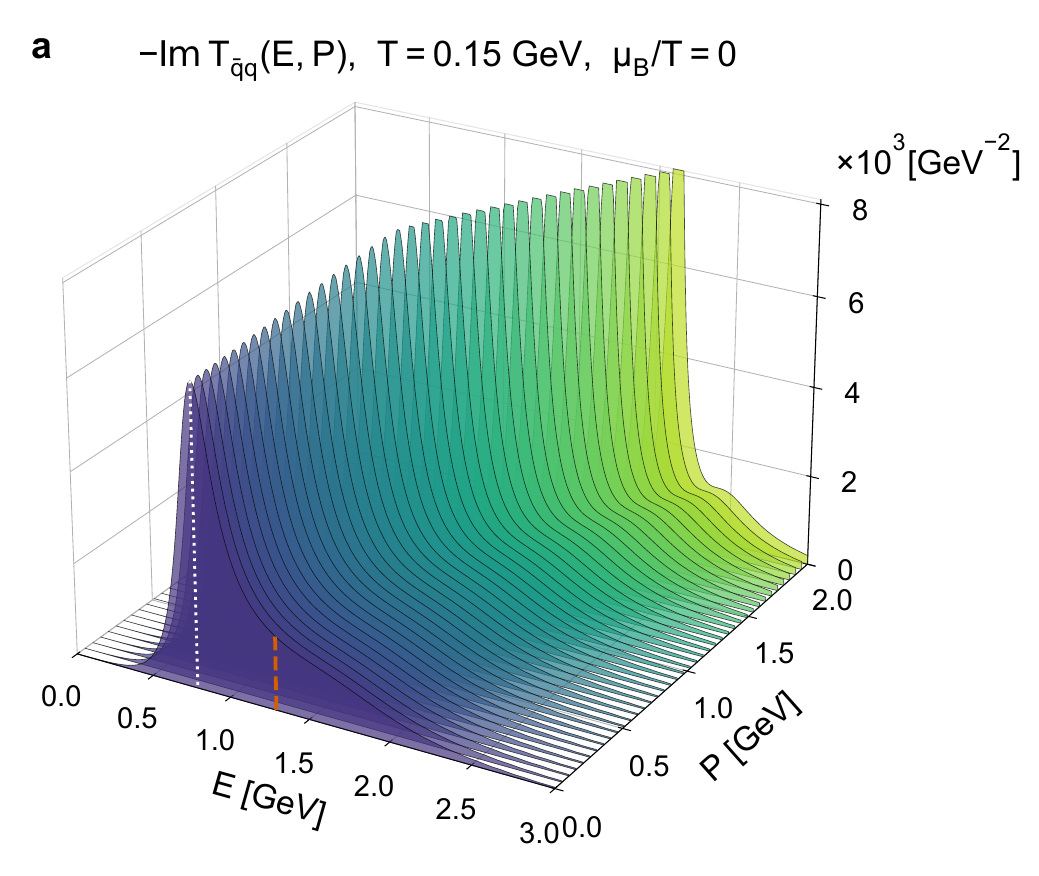}
	\includegraphics[width=.45\linewidth]{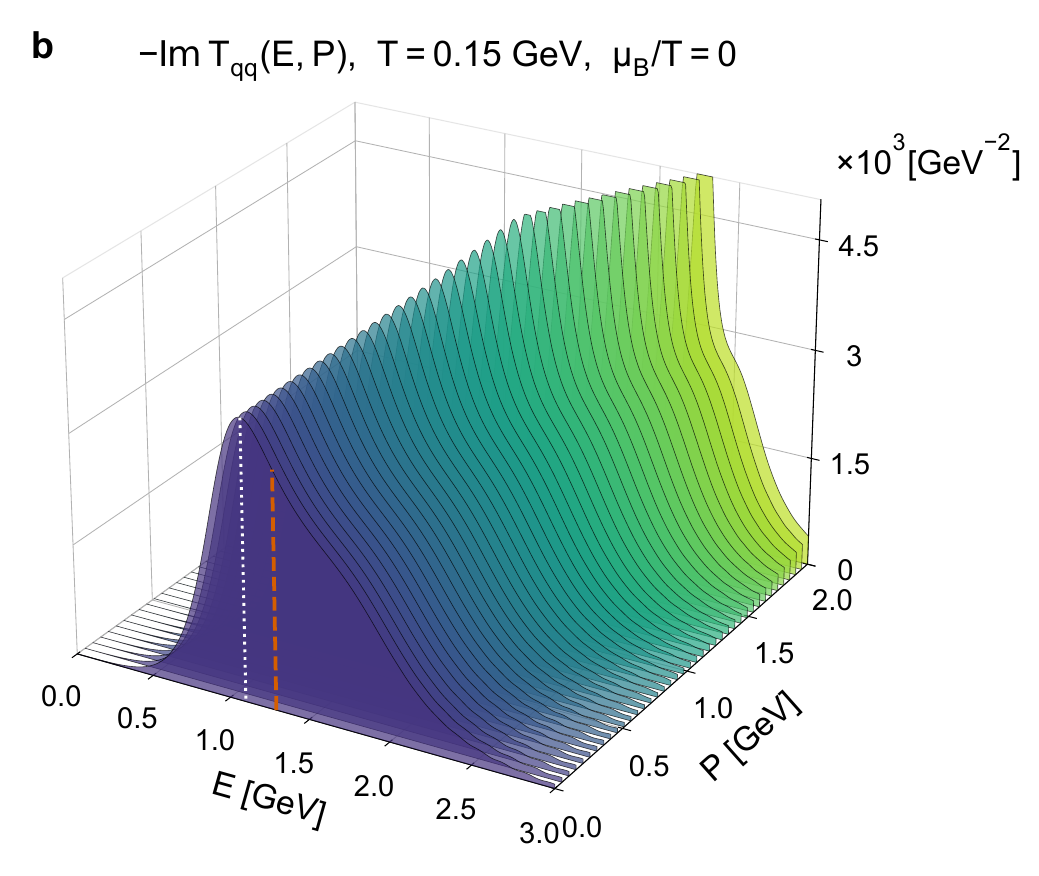}
	\caption{\textbf{Imaginary parts of the $T$--matrices in the QGP.} $-\Im T(E, P)$ at $T = 0.15\gev$ and $\mu = 0$ for the \textbf{a,} meson channel $q\bar q$  and \textbf{b,} diquark channel $qq$, drawn as filled ridges, one energy slice per pair momentum $P$. Dashed vertical lines indicate the 2-body threshold $E=2m_q$ and dotted vertical lines mark the peaks of $-\Im T$ at $P=0$, which determine the masses of the corresponding resonances.}
	\label{fig::ext::Tmat_mu0}
\end{figure*}

\begin{figure*}
	\includegraphics[width=.45\linewidth]{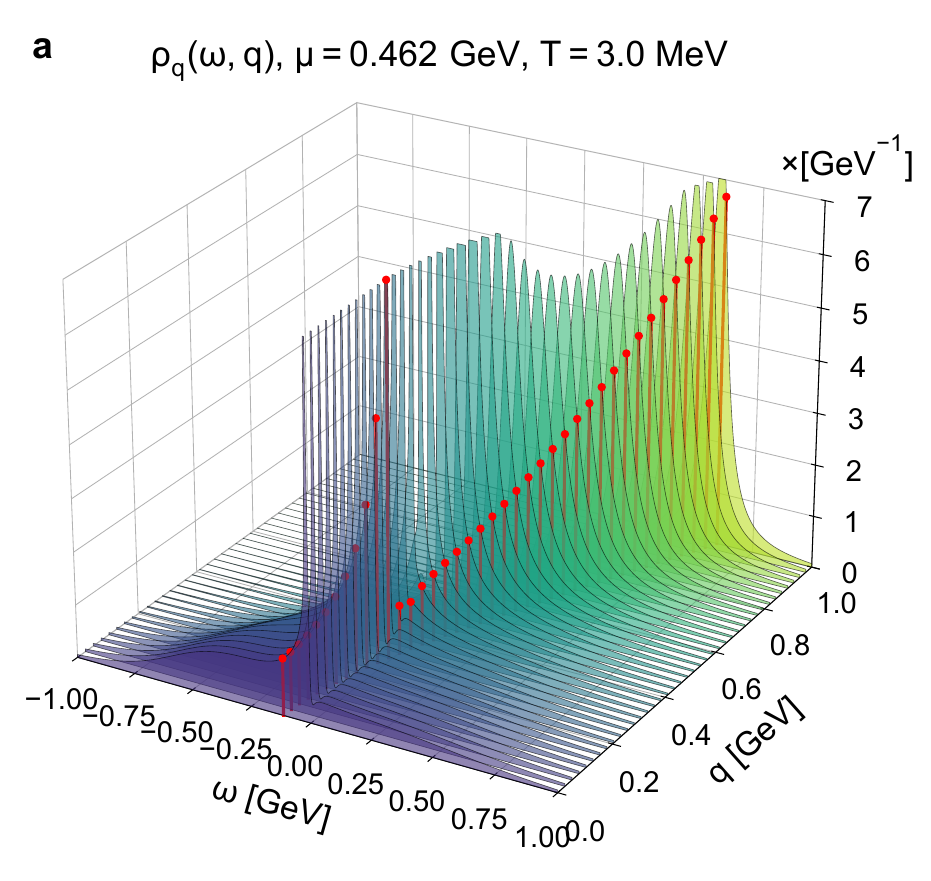}
	\includegraphics[width=.45\linewidth]{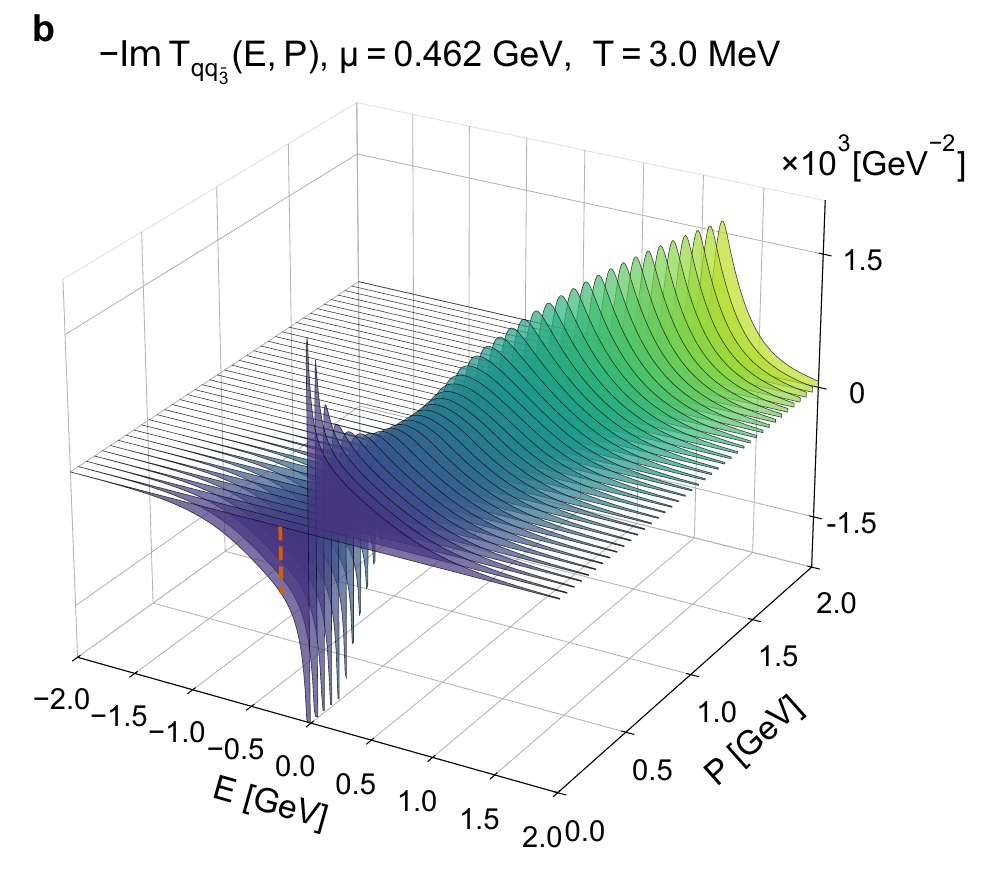}

	\caption{\textbf{Spectral properties in the cold dense regime at $T = 3\mev$ and $\mu = 0.462$~GeV}. \textbf{a,} Self-consistent quark spectral function $\rho_q(\omega, q)$, with frequencies measured relative to the Fermi level, drawn as a stack of ridges, one frequency slice per momentum $q$. Vertical lines with dots show the QP dispersion relation $\omega = \omega_q(q) - \mu$. \textbf{b,} Imaginary part of the self-consistent $T$-matrix in the antitriplet quark-quark channel, $-\Im T_{qq_{\bar 3}}(E, P)$, as a function of the pair energy $E$ and pair momentum $P$. The dashed line on the $P=0$ slice denotes a QP 2-body threshold $2 m_q - 2 \mu$.}
	\label{fig::ext::lowT}
\end{figure*}

\begin{table*}[t!]
\centering
\begin{minipage}[t]{0.56\linewidth}
\centering
\caption{Parameters of the rank-3 separable interaction (in powers of GeV where dimensionful).}
\label{tab:params}
\begin{tabular}{lr|lr|lr}
\hline\hline
\multicolumn{2}{c|}{$v_1$ (dipole)} & \multicolumn{2}{c|}{$v_2$ (Gaussian)} & \multicolumn{2}{c}{$v_3$ (shifted dipole)} \\
\hline
$|G_1|$          & 8.834 & $|G_2|$          & 0.354 & $|G_3|$          & 0.876 \\
$\Lambda_1$      & 0.537 & $\Lambda_2$      & 1.500 & $\Lambda_3$      & 1.140 \\
$S_1$            & 0.594 & $C_2$            & 0.048 & $q_0$            & 1.500 \\
$\alpha_1^{(p)}$ & 1.103 & $\alpha_2^{(p)}$ & 1.609 & $q_0^{(p)}$      & 2.128 \\
                 &       &                  &       & $\alpha_3^{(p)}$ & 0.384 \\
\hline
\multicolumn{2}{c|}{Screening} & \multicolumn{2}{c|}{GZ gluon} & \multicolumn{2}{c}{Mass ansatz} \\
\hline
$W$              & 0.150 & $m_g$            & 1.200 & $\alpha_1$       & 2.0 \\
$T_0$            & 0.150 & $\gamma$         & 0.850 & $\alpha_2$       & 4.0 \\
$\alpha_1^v$     & 2.0   &                  &       & $\gamma_2$ & -0.15 \\
$\alpha_2^v$     & 4.0   &                  &       &                  & \\
\hline\hline
\end{tabular}
\end{minipage}\hfill
\begin{minipage}[t]{0.40\linewidth}
\centering
\caption{Casimir factors and degeneracies $({\cal F}_a,\, d_a)$ for the two-body color channels.
Attractive (repulsive) channels have ${\cal F}_a > 0$ ($< 0$).}
\label{tab:casimir}
\begin{tabular}{ccccc}
\hline\hline
 & $qq$ & $q\bar q$ & $(q/\bar q)\, g$ & $gg$ \\[2pt]
\hline\\[-8pt]
 & $(\frac{1}{2},\, \bar{\mathbf 3})$ & $(1,\, \mathbf{1})$ & $(\frac{9}{8},\, \mathbf{3})$ & $(\frac{9}{4},\, \mathbf{1})$ \\[4pt]
 & $(-\frac{1}{4},\, \mathbf{6})$ & $(-\frac{1}{8},\, \mathbf{8})$ & $(\frac{3}{8},\, \mathbf{6})$ & $(\frac{9}{8},\, \mathbf{16})$ \\[4pt]
 & & & $(-\frac{3}{8},\, \mathbf{15})$ & $(-\frac{3}{4},\, \mathbf{27})$ \\[4pt]
\hline\hline
\end{tabular}
\end{minipage}
\end{table*}

\end{document}